\documentclass[reqno]{amsart}
\usepackage{amsmath,amssymb,cite}
\usepackage[mathscr]{euscript}
\theoremstyle{plain}

\theoremstyle{definition}

\theoremstyle{remark}

\numberwithin{equation}{section}
\numberwithin{theorem}{section}
\newcommand{\mc}[1]{{\mathcal #1}}
\newcommand{\mb}[1]{{\mathbf #1}}
\newcommand{\bb}[1]{{\mathbb #1}}

\newcommand{\upbar}[1]{\,\overline{\! #1}}

\newcommand{\id}{{1 \mskip -5mu {\rm I}}}
\renewcommand{\epsilon}{\varepsilon}

\newcommand{\lan}{\langle}
\newcommand{\ran}{\rangle}

\title[Drift of phase fluctuations in the ABC model]
{Drift of phase fluctuations \\ in the ABC model}

\author[L.\ Bertini]{Lorenzo Bertini}
\address{Lorenzo Bertini \hfill\break \indent
   Dipartimento di Matematica, Sapienza Universit\`a di Roma
   \hfill\break \indent
   P.le Aldo Moro 5, 00185 Roma, Italy}
 \email{bertini@mat.uniroma1.it}

\author[P.\ Butt\`a]{Paolo Butt\`a}
\address{Paolo Butt\`a\hfill\break \indent
   Dipartimento di Matematica, Sapienza Universit\`a di Roma 
   \hfill\break \indent
   P.le Aldo Moro 5, 00185 Roma, Italy}
 \email{butta@mat.uniroma1.it}

\begin{document}

\begin{abstract}
In a recent work, Bodineau and Derrida analyzed the phase fluctuations in the ABC model. In particular, they computed the asymptotic variance and, on the  basis of numerical simulations, they conjectured the presence of  a drift, which they guessed to be an antisymmetric function of the three densities. By assuming the validity of the fluctuating hydrodynamic approximation, we prove the presence of such a drift, providing an analytical expression for it. This expression is then shown to be an antisymmetric function of the three densities. The antisymmetry of the drift can also be inferred from a symmetry property of the underlying microscopic dynamics. 
\end{abstract}


\maketitle
\thispagestyle{empty}

\section{Introduction}
\label{s:i}

The \emph{ABC model}, introduced by Evans et al.\ \cite{5,6}, is a
one-dimensional stochastic conservative dynamics with local jump rates,
whose invariant measure undergoes a phase transition.  It is a system
consisting of three species of particles, traditionally labeled $A$,
$B$, and $C$, on a discrete ring with $L$ sites. The system evolves by
nearest neighbor particles exchanges with the following 
rates: $AB \to BA$, $BC \to CB$, $CA \to AC$ with rate $q$
and $BA \to AB$, $CB \to BC$, $AC \to CA$ with rate $1/q$.  In
particular, the total number of particles $N_\alpha$, of each species 
$\alpha\in\{A,B,C\}$,  are conserved and satisfy $N_A+N_B+N_C=L$. 
When $q\neq 1$, Evans et al.\ \cite{5,6} argued that in the thermodynamic
limit $L \to \infty$ with $N_\alpha /L \to r_\alpha$ the system
segregates into pure $A$, $B$, and $C$ regions, with translationally
invariant distribution of the phase boundaries.  In the equal
densities case $N_A=N_B=N_C=L/3$ the dynamics is reversible and its
invariant measure can be explicitly computed.  As shown in
\cite{FF1,FF2}, the ABC model can be reformulated in terms of a dynamic
of random walks on the triangular lattice.

As discussed by Clincy et al.\ \cite{8}, the natural scaling to
investigate the asymptotic behavior of the ABC model is the
\emph{weakly asymmetric} regime $q=\exp\big\{-\frac{\beta}{2L}\big\}$,
where the parameter $\beta$ plays the role of an
inverse temperature.  With this choice, the reversible measure of the
equal densities case $r_A=r_B=r_C=1/3$ becomes a canonical Gibbs
measure with a mean field Hamiltonian, which undergoes a second order
phase transition at $\beta_{\mathrm{c}}= 2\pi\sqrt{3}$.
More precisely, for $\beta \le \beta_\mathrm{c}$ 
the typical densities profiles are homogeneous while  
for  $\beta>\beta_\mathrm{c}$ the three species segregate. 
For unequal densities the invariant measure of the ABC dynamics on a
ring is not reversible and cannot be computed explicitly. 
The asymptotic of the two-point correlation functions in the
homogeneous phase is obtained in \cite{bdlw,8}, where 
the large deviation rate function for the stationary measure 
is also calculated up to order $\beta^2$.
When the ABC dynamics is considered on an open interval with zero flux
condition at the endpoints, the corresponding invariant measure is
reversible for all values of the densities \cite{Ayetal}.  In
particular, it has the same Gibbs form as the one in the ring for the
equal density case. In this paper we shall however stick to the case
of periodic boundary conditions.

\noindent\emph{Diffusive scaling limit.} The hydrodynamic behavior of the
empirical densities $\rho =\rho(x,t)= \big(\rho_A(x,t),\rho_B(x,t),
\rho_C(x,t)\big)$, where $t\ge 0 $ and $x\in \mb T$, the one-dimensional 
torus of length one, is obtained by a diffusive rescaling by space and time. In this limit the empirical density evolves according to the deterministic parabolic system,
\begin{equation}
  \label{hy}
  \begin{split}
    \frac{\partial \rho_A}{\partial t} & = \frac{\partial^2 \rho_A}{\partial x^2} 
    +\beta \, \frac{\partial}{\partial x}\big[ \rho_A (\rho_B-\rho_C)\big],
    \\
     \frac{\partial \rho_B}{\partial t} & = \frac{\partial^2 \rho_B}{\partial x^2}
    +\beta \, \frac{\partial}{\partial x}\big[ \rho_B (\rho_C-\rho_A)\big],
    \\
     \frac{\partial \rho_C}{\partial t} & = \frac{\partial^2 \rho_C}{\partial x^2}
    +\beta \, \frac{\partial}{\partial x}\big[ \rho_C (\rho_A-\rho_B)\big].
  \end{split}
\end{equation}
The proof of such a statement could be achieved by using standard tools in hydrodynamical limits, see e.g., \cite{KL,S}; see also \cite{FF3} for an alternative method. The initial conditions for the system \eqref{hy} are determined by the starting microscopic configuration, they satisfy the constraints $\rho_A(\cdot,0)+\rho_B(\cdot,0)+\rho_C(\cdot,0)=1$ and 
$0\le \rho_\alpha(\cdot,0) \le 1$, $\alpha\in\{A,B,C\}$, 
which are preserved by the above flow. Clearly, it also preserves the
mass of each species, i.e., $r_\alpha = \int \!\mathrm{d}x\,
\rho_\alpha(x,t)$, $\alpha\in\{A,B,C\}$, is constant in time.  
We mention that in the equal densities case, $r_A=r_B=r_C=1/3$, the
flow defined by \eqref{hy} is a (suitable) \emph{gradient flow} of the
large deviations rate function, which in this case reads,
\begin{equation}
  \label{Fhy}
\mc F_\beta(\rho) = \int_0^1\!\mathrm{d}x\, \sum_{\alpha}\rho_\alpha(x) \log\rho_\alpha(x) + \beta \int_0^1\!\mathrm{d}x\int_x^1\!\mathrm{d}y\:\sum_\alpha\rho_\alpha(x)\rho_{\alpha+2}(y),
\end{equation}
where the sum on the species label $\alpha$ is modulo three. This is the macroscopic
counterpart of the reversibility of the underlying microscopic dynamics.

In the equal density case the stationary solutions to \eqref{hy} correspond to the critical points of \eqref{Fhy}, which have been analyzed in \cite{Ayetal}. In particular, the homogenous profile $r=(\frac 13, \frac 13, \frac 13)$ is the unique stationary solutions for $\beta\in [0,\beta_\mathrm{c}]$, while for $\beta>\beta_\mathrm{c}$ there exists a one-periodic inhomogeneus stationary solution, unique up to translations, which minimizes the large deviation functional and it is therefore stable for the flow \eqref{hy}. 
In the general case, the homogeneous profile $r=(r_A,r_B,r_C)$ is clearly still a stationary
solution to \eqref{hy}. For $\beta$ small enough (depending on $r$) it is the unique
one. As discussed in \cite{bdlw,8}, a linear stability analysis shows that for
$\beta>\beta_{r} = 2\pi/\sqrt{1 -2 r^2}$, $r^2=r_A^2+r_B^2+r_C^2$, it becomes unstable.  
As stated there, the phase transition, at least for particular values
of $r$, is expected to become of first order. 
Namely, for some $\beta \in (0,\beta_r)$ there exist other stationary
solutions to \eqref{hy} which actually describe the typical profiles with respect to
the invariant measure of the underlying microscopic dynamics.  
In \cite{CM} the stationary solutions to \eqref{hy} are characterized
and their explicit expression is written in terms of elliptic functions (see also \cite{FF1},
where similar results were obtained for some general exclusion processes, including ABC with not necessarily weakly asymmetric transition rates).  
This analysis reveals that for some $r$ and $\beta \in (0,\beta_r)$ there 
exist one-periodic stationary solutions to \eqref{hy},
in agreement with the conjectured occurrence of a first order transition. 

The next natural issue on \eqref{hy} is whether it admits traveling
wave solutions, this is indeed a side question both in \cite{BD} and
\cite{CM}. A simple argument, discussed in \cite{Q} shows that such solutions do 
not occur. For the reader's convenience we report the argument in 
Appendix~\ref{s:ntw}. We emphasize that the periodicity of the space variable is crucial.  Indeed, when the system \eqref{hy} is considered on the whole line it does admit traveling waves. 

\noindent\emph{Beyond the diffusive scaling.} The main result of the present paper is the identification of the drift for the phase fluctuations. In order to describe it, fix values
of $r$ and $\beta$ in the low temperature part of the phase diagram.
The analysis in \cite{CM} implies the existence of a one-periodic
profile $\upbar\rho=\big(\upbar{\rho}_A,\upbar{\rho}_B,\upbar{\rho}_C\big)$
such that the one-periodic stationary solutions to \eqref{hy} are
given by the translations of $\upbar\rho$. Observe that, as discussed
in \cite{Ayetal,CM}, for larger values of $\beta$ there exist also
$\frac 12$-periodic solutions, $\frac 13$-periodic solutions,...  However, as
proven in \cite{Ayetal} in the case of equal densities and suggested
by numerical evidences in general, these less segregated profiles are
not expected to describe the typical behavior of the microscopic
evolution.  Consider now the microscopic dynamics in which the initial
distribution of the species is associated to the macroscopic profile
$\upbar\rho$. The hydrodynamical description discussed above implies
that on the microscopic time scale $O(L^2)$ the associated
macroscopic profile does not move. Since macroscopic fluctuations are
$O(L^{-\frac 12})$, by taking into account the translation invariance, at
times $O(L^3)$ the macroscopic profile is expected to perform a random
motion on the set $\big\{\upbar{\rho}(\cdot-z), \, z\in \mb T\big\}$.
This motion is refereed to as \emph{phase fluctuations}.  In agreement
with the above dynamical picture, we observe that in the equal
densities case, by sampling the particles according to the Gibbsian
invariant measure, the law of large numbers for the empirical
densities of the three species is $\upbar{\rho}(\cdot-\zeta)$, where
$\zeta$ is a uniform random variable on $\mb T$, see
\cite{BCP} for a formal proof of this statement. 

Bodineau and Derrida \cite{BD} have computed the variance of the
phase fluctuations by using both the methods of the fluctuating
hydrodynamics and of the macroscopic fluctuation theory.  Moreover, on
the basis of numerical evidences, further confirmed in \cite{CM}, they
conjectured the presence of a drift which they guessed to be an
antisymmetric function of the three densities. Observe that for equal densities, 
reversibility readily implies that the drift vanishes. In this paper, by
assuming the validity of the fluctuating hydrodynamic approximation for
the ABC model, we prove in general the presence of such a drift. In particular, we deduce an
analytical expression for the drift, $v=v(\beta;r_A,r_B,r_C)$, see equation \eqref{p7} below, in terms of the semigroup generated by the linearization of the hydrodynamic equations around $\upbar{\rho}$. We also show that, as conjectured in \cite{CM}, $v(\beta;r_A,r_B,r_C)$ is antisymmetric with respect to the exchange of the masses $r_A,r_B,r_C$. We finally analyze the behavior of the drift near the second order phase transition, showing that, in contrast to the variance, it does not diverge. In this respect, we also quote \cite{GD1,GD2}, where current fluctuations and long-range correlations at the phase transition are analyzed.

In order to obtain the statistics of the phase fluctuations, we need to obtain an effective dynamics on the manifold $\big\{\upbar{\rho}(\cdot-z), \, z\in \mb T\big\}$.  The fluctuating hydrodynamic approximation allows to describe the behavior of the microscopic ABC dynamics by the hydrodynamic equations \eqref{hy} perturbed with an additive noise of order $L^{-\frac 12}$. In particular, the behavior of the ABC model on the time scale $O(L^3)$ is captured by the asymptotics of the fluctuating hydrodynamics on the time scale $O(L)$. Under this assumption, the variance of the phase fluctuations can be computed by considering the projection along the manifold of the noise: this is the sum of  independent, mean zero, order $L^{-\frac 12}$ random variables, that converges to a Brownian motion. On the other hand, the origin of the drift is much less evident and due to the nonlinearity of \eqref{hy}. Indeed, we show that in a time step $T$ with $1\ll T \ll L$ the nonlinear term gives a deterministic contribution to the effective dynamics of order $T/L$, which sum up to a finite drift at times $O(L)$. 

We finally mention that the statistics of dynamical phase fluctuations has been already analyzed in the context of the one-dimensional nonconservative stochastic Ginzburg-Landau equation \cite{BBDP,BDP,F}. In particular, if the reaction term is not symmetric, the mechanism outlined above gives rise to a constant drift in the resulting random motion. This has been rigorously proved in \cite{BB}. Phase fluctuations for the Kuramoto model, which is a mean field conservative dynamics, have been recently discussed in \cite{BGP}.
 
\section{Identification of the drift}
\label{s:1}

In this section we introduce the fluctuating hydrodynamic assumption
and identify the drift of the phase fluctuations in terms of the
semigroup generated by the linearization of \eqref{hy} around the
periodic profile $\upbar\rho$. We shall frequently refer to
\cite{BD} and use the same notation introduced there. For convenience
we set $\epsilon=1/L$.

\subsection{The fluctuating hydrodynamic assumption}

At the macroscopic level, the effect of the microscopic fluctuation
can be modeled by adding to the hydrodynamic system
\eqref{hy} a suitable random force whose statistics can be inferred by 
an informal computation on the underlying Markov dynamics. 
Referring to either Appendix \ref{s:md} or \cite{BD,bdlw,8} for the details of such computation, the
corresponding stochastic system reads, using vector notation,
\begin{equation}
  \label{fhy}
\frac{\partial}{\partial t} 
  \begin{pmatrix}
    \rho_A\\ 
    \rho_B\\ 
    \rho_C
  \end{pmatrix}
  = 
 \frac{\partial^2}{\partial x^2}  
 \begin{pmatrix}
     \rho_A\\ 
    \rho_B\\ 
    \rho_C
   \end{pmatrix}
    + 
  \beta \frac{\partial}{\partial x}  
  \begin{pmatrix}
      \rho_A(\rho_B-\rho_C) \\ 
    \rho_B(\rho_C-\rho_A) \\ 
    \rho_C(\rho_A-\rho_B) 
  \end{pmatrix}
  + \sqrt{\epsilon}  
\frac{\partial}{\partial x}  
\begin{pmatrix}    
\eta_A^\epsilon\\ 
    \eta_B^\epsilon\\ 
    \eta_C^\epsilon
  \end{pmatrix}
\end{equation}
where $t\ge 0$, $x\in \mb T=\mb R/\mb Z$, the one-dimensional torus, and, conditionally on the value $\rho$, the noise $\eta^\epsilon=(\eta_A^\epsilon,\eta_B^\epsilon,\eta_C^\epsilon)$ is Gaussian with correlations
\begin{equation}
  \label{cov0}
  \big\langle \eta^\epsilon_\alpha(x,t) \eta^\epsilon_{\alpha'} (x',t')
  \big\rangle
  = \Sigma_{\alpha,\alpha'} (\rho;x,x') \, \delta_\epsilon(x-x') \, \delta(t-t'), \qquad \alpha,\alpha'\in\{A,B,C\}.
\end{equation}
Above, $\delta_\epsilon$ is the $\epsilon$-approximation to the Dirac's $\delta$-function, i.e., its length  scale is of order $\epsilon$, and $\Sigma$ is the matrix
\begin{equation}
\label{p1.5-1}
\Sigma(\rho;x,x') =  D(\rho;x,x') + D(\rho;x',x)
\end{equation}
in which $D(\rho;x,x')$ is
\begin{equation*}  
\begin{pmatrix} 
    \rho_A(x)[\rho_B(x')+\rho_C(x')] & -\rho_A(x)\rho_B(x') & -\rho_A(x)\rho_C(x') \\
    -\rho_A(x)\rho_B(x') &   \rho_B(x)[\rho_A(x')+\rho_C(x')] & -\rho_B(x)\rho_C(x') \\ 
    -\rho_A(x)\rho_C (x')& -\rho_B(x)\rho_C(x') &  \rho_C(x)[\rho_A(x')+\rho_B(x')]
  \end{pmatrix}.   
\end{equation*}
We observe that $\Sigma$ has a vanishing eingenvalue whose corresponding eigenspace is spanned by the vector $(1,1,1)^\top$. Accordingly, \eqref{fhy} preserves the constraint $\sum_\alpha \rho_\alpha =1$. 

Clearly, the (deterministic) hydrodynamic system \eqref{hy} is
recovered from \eqref{fhy} simply by setting $\epsilon=0$. As
discussed e.g., in \cite[\S~II.3.5]{S}, \eqref{fhy} also
predicts the nonequilibrium Gaussian fluctuations (in the diffusive scaling limit) which can be inferred by linearizing \eqref{fhy} around an hydrodynamic solution. As discussed in the Introduction, in our analysis we shall need the stronger assumption that \eqref{fhy} 
correctly encodes the behavior of the ABC dynamics on time scales
longer then the hydrodynamical one. 
More precisely, as phase fluctuations becomes observable (i.e., macroscopically of
order one) at microscopic times of order $L^3$, the connection between
the analysis on the stochastic system \eqref{fhy} performed in the sequel and 
the microscopic dynamics relies on the
hypotheses that the ABC dynamics at times $L^3$ can be captured by
looking at \eqref{fhy} on times of order $\epsilon^{-1}$ and then
taking the limit $\epsilon\to 0$.

In regard to such fluctuating hydrodynamic assumption, we simply observe that the use of nonlinear stochastic equation to describe the evolution of particle systems beyond the hydrodynamic scale is a common practice in nonequilibrium statistical physics. On the other hand, the rigorous justification of such procedure is a most challenging task of mathematical physics, see \cite{BG} for the weakly asymmetric exclusion process which has, albeit much simpler, similar feature to the ABC model. We finally emphasize that in \eqref{fhy} we have somehow kept track of the underlying discrete structure by using a colored noise with spatial correlation length of order $\epsilon$. With this choice, we avoid in particular the difficult problem of giving a precise mathematical meaning to \eqref{fhy} with a space-time white noise, see \cite{H} for the KPZ equation whose nonlinearity has a similar structure. As it will clear in the computation of the drift, the problem of ultraviolet singularities will however appear when taking the limit $\epsilon\to 0$. 

\subsection{The effective phase dynamics}

It is convenient to take advantage of the constraint $\rho_A+\rho_B+\rho_C=1$ and write the system \eqref{fhy} in terms of 
\[
\rho = \rho(x,t) = \begin{pmatrix} \rho_A(x,t) \\ \rho_B(x,t) \end{pmatrix}.
\]
Denoting by  
\begin{equation}
\label{N}
\mc N(\rho) = \begin{pmatrix} \rho_A^2+2\rho_A\rho_B -\rho_A \\
-\rho_B^2-2\rho_A\rho_B +\rho_B \end{pmatrix},\qquad \eta^\epsilon = \begin{pmatrix} \eta_A^\epsilon \\
\eta_B^\epsilon \end{pmatrix}
\end{equation}
the nonlinear transport and the noise, respectively, we thus rewrite \eqref{fhy} as 
\begin{equation}
\label{p1}
\frac{\partial\rho}{\partial t} = \frac{\partial^2\rho}{\partial x^2}  + \beta \, \frac{\partial}{\partial x}  \mc N(\rho) +\sqrt\epsilon\,\frac{\partial\eta^\epsilon}{\partial x}. 
\end{equation}
Conditionally on the value $\rho$, the noise $\eta^\epsilon$ is Gaussian with correlations as in \eqref{cov0}, where $\Sigma(\rho;x,x')$ is the $2\times 2$ matrix as in \eqref{p1.5-1} in which now
\begin{equation}
\label{p1.5}
D(\rho;x,x') = 
 \begin{pmatrix} 
    \rho_A(x)[1-\rho_A(x')] & -\rho_A(x)\rho_B(x') \\
    -\rho_A(x)\rho_B(x') &   \rho_B(x)[1-\rho_B(x')] 
  \end{pmatrix}.   
\end{equation}

Hereafter, we fix the total densities $r=(r_A,r_B)$ and $\beta>0$ in the low temperature part of the phase diagram established in \cite{CM}. The analysis therein implies the existence of a one-periodic profile
\[
\upbar{\rho} = \upbar{\rho} (x) =  \begin{pmatrix} \upbar{\rho}_A(x) \\ \upbar{\rho}_B(x) \end{pmatrix}
\]
which solves the following system of ordinary differential equation,
\begin{equation}
  \label{ss}
\upbar{\rho}'' +\beta  \mc N (\upbar{\rho})' =0,
\end{equation}
and satisfies the constraints $\upbar{\rho}_\alpha \ge 0$, $\upbar{\rho}_A + \upbar{\rho}_B \le 1$, and $\int\!\mathrm{d} x\, \upbar{\rho}_\alpha = r_\alpha$, $\alpha\in \{A,B\}$. The one-periodic stationary solutions to \eqref{p1} with $\epsilon=0$ are given by the translations of $\upbar\rho$. 

We denote by $\upbar{\rho}_z(x)=\upbar{\rho}(x-z)$ the translation of $\upbar{\rho}$ by $z$ on the torus. Let also $\mc L_z$ be the linear part of \eqref{p1} around
$\upbar{\rho}_z$, i.e., 
\begin{equation}
\label{p2}
\mc L_z\psi = \frac{\partial^2\psi}{\partial x^2} + \beta \frac{\partial}{\partial x} (\mc B_z\psi),
\end{equation}
where
\begin{equation}
\label{p3}
\mc B_z(x)=\mc B(x-z), \qquad \mc B =\begin{pmatrix} 2\upbar{\rho}_A+2\upbar{\rho}_B-1 & 2\upbar{\rho}_A \\ -2\upbar{\rho}_B & -2\upbar{\rho}_A-2\upbar{\rho}_B+1 \end{pmatrix}.  
\end{equation} 
Since the nonlinear evolution preserves the masses, we regard $\mc L_z$ as an operator on the space of mean zero function. In view of \cite{BD},
\begin{equation}
\label{p35}
\chi_z(x) = \big( \upbar{\rho}_B(x-z)-r_B,r_A -\upbar{\rho}_A(x-z) \big), \qquad
\upbar{\rho}_z'(x) = \begin{pmatrix} \upbar{\rho}_A'(x-z) \\ \upbar{\rho}_B'(x-z) \end{pmatrix}
\end{equation}
are the left and right eigenvectors of $\mc L_z$ with zero eigenvalue, which can be easily shown to be a simple eigenvalue. The rest of the spectrum is composed by a countable set of eigenvalues. As in \cite{BD}, we assume they have strictly negative real part, bounded away from zero. In regard to this assumption, apart from the numerical evidence, see e.g., \cite{CM}, we remark that it can be verified analytically in two regimes. The first is when the total densities $(r_A,r_B,r_C)$ are close to $(\frac 13,\frac 13,\frac 13)$. In the equal density case $\mc L_z$ can be realized as a nonnegative self-adjoint operator on a suitable Hilbert space, see \cite[Remark 4.1]{BD}, and the above assumption is fulfilled. A standard perturbation argument yields the statement. The second regime is when $\beta$ and $r$ are close to the second order phase transition. The result follows by Appendix \ref{app:c}, where the operator $\mc L_z$ is analyzed as a perturbation of the differential operator with constant coefficients obtained by linearizing \eqref{hy} around the homogeneous profile. 

Under the above assumption, the manifold $\mc M = \{\upbar{\rho}_z\colon z\in \mb T\}$ is locally exponentially attractive for the deterministic flow \eqref{hy}. The projection onto the null space of $\mc L_z$ is given by the tensor product $\gamma\upbar{\rho}_z'(x)\chi_z(x')$, where 
\begin{equation}
\label{gamma}
\frac 1\gamma = \int_0^1\!\mathrm{d} x\: \chi\,\upbar{\rho}'. 
\end{equation}
We shall denote by $\mc P_z$ the projector whose integral kernel is given by $\mc P_z(x,x')= \delta(x-x')\id - \gamma\upbar{\rho}_z'(x)\chi_z(x')$. Observe also that the right eigenvector $\upbar{\rho}_z'$ is the infinitesimal generator of the translations on $\mc M$.

We consider the random flow \eqref{fhy} with an initial condition lying in a $\epsilon^{\frac 12}$-neighbor\-hood of $\mc M$. As fluctuations transversal to $\mc M$ are exponentially damped by the deterministic part of the flow, we deduce that the solution remains in such neighborhood until a large fluctuation takes place. For our purposes, those large fluctuations can be neglected, as their probability is exponentially small in $\epsilon^{-1}$ up to time scales polynomially large in $\epsilon^{-1}$. Indeed, such a fluctuation requires a very large deviation of the noise to overcame the deterministic flow. We conclude that, with probability close to one, the solution to \eqref{fhy} remains in a $\epsilon^{\frac 12}$-neighborhood of $\mc M$ up to the time scale of order $\epsilon^{-1}$, which is the relevant one for the phase fluctuations.

In order to describe the motion along the manifold $\mc M$ we use the Fermi coordinates $(\zeta,\psi)$, defined as it follows. Given $\rho$ in a neighborhood of $\mc M$, the angular coordinate $\zeta\in \mb T$, called the \emph{center} of $\rho$, is defined as the point $z$ such that the component of $\rho - \upbar{\rho}_z$ in the null space of $\mc L_z$ vanishes, namely, the solution to 
\begin{equation}
\label{p3.5}
F(z) = \int_0^1\!\mathrm{d} x\: \chi_z \, (\rho-\upbar{\rho}_z) = 0
\end{equation}
and then we set $\psi=\rho-\upbar{\rho}_\zeta$, so that $\psi = \mc P_\zeta \psi$. An application of the implicit function theorem shows that $z$ is uniquely defined if $\rho$ lies in a small neighborhood of $\mc M$.

According to the scheme introduced in \cite{BDP} and further developed in \cite{BBDP}, the motion along the manifold $\mc M$ can be identified by the following recursive procedure. Given an initial datum $\rho(\cdot,0)$ in a $\epsilon^{\frac 12}$-neighborhood of $\mc M$, we let $\zeta_0$ be its center and decompose $\rho(\cdot,0)=\upbar{\rho}_{\zeta_0}(\cdot)+\psi(\cdot,0)$. We then linearize the evolution \eqref{p1} around $\upbar{\rho}_{\zeta_0}$ and  compute the displacement in a time interval $1\ll T \ll \epsilon^{-1}$. At this point, we recenter the solution by computing the new center $\zeta_T$ at time $T$ and then iterate. As it will be clearer in the sequel, after $(\epsilon T)^{-1}$ steps we get a finite displacement of the center, which has the form of a Brownian motion with a constant drift on $\mb T$. Referring to e.g., \cite{BDP}, for the precise mathematical construction in terms of stopped martingales, we next detail the first step of such procedure in which we drop the subscript $\zeta_0$ from the notation. By introducing $u=u(x,t)$ as
\[
u = \begin{pmatrix}u_A \\u_B \end{pmatrix} = \rho-\upbar{\rho},
\]
the evolution \eqref{p1} can be recast into the form,      
\begin{equation}
\label{p4}
\frac{\partial u}{\partial t} = \mc L u + \beta \frac{\partial}{\partial x}\mc N_1(u)  + 
\sqrt\epsilon\, \frac{\partial \bar\eta^\epsilon}{\partial x} + \cdots , 
\end{equation}
where
\begin{equation}
\label{p4.25}
\mc N_1(u) = \begin{pmatrix} u_A^2+2u_Au_B \\ -u_B^2-2u_Au_B \end{pmatrix}, 
\end{equation}
the operator $\mc L$ is as in \eqref{p2}-\eqref{p3}, and we approximated $\eta^\epsilon$ with the Gaussian noise $\bar\eta^\epsilon$, having covariance matrix $\Sigma(\upbar{\rho};x,x')=D(\upbar{\rho};x,x')+D(\upbar{\rho};x',x)$, with $D(\rho;x,x')$ as in  \eqref{p1.5}. Note indeed that in a single step of the iteration the solution remains close to the initial condition and, as it will be clearer in the sequel, the approximation in \eqref{p4} does not affect the phase fluctuations. By Duhamel formula,  
\begin{equation}
\label{p5}
u(t) = \mathrm{e}^{t\mc L}\psi(0) + \beta \int_0^t\!\mathrm{d} s\,
\mathrm{e}^{(t-s)\mc L} \frac{\partial}{\partial x}\mc N_1(u(s)) + \sqrt\epsilon\, W^\epsilon(t) +
\cdots, 
\end{equation}
where $W^\epsilon(t) = \begin{pmatrix} W^\epsilon_A(t) \\ W^\epsilon_B(t) \end{pmatrix}$ is Gaussian with covariance
\begin{equation}
\begin{split}
\label{p5.5}
& \left\langle W^\epsilon(x,t)\, W^\epsilon(x',t)^\top \right\rangle \\ & \quad = \int_0^t\!\mathrm{d} s \! \int_0^1\!\mathrm{d} y\! \int_0^1\!\mathrm{d} y' \frac{\partial G}{\partial y}(x,y,s) \delta_\epsilon(y-y')  \Sigma(\upbar{\rho};y,y')\: \frac{\partial G}{\partial y}(x',y',s)^\top  
\end{split}
\end{equation}
and $G(x,y,t) = \mathrm{e}^{t\mc L}(x,y)$ is the fundamental solution associated to $\mc L$. 

Since $\psi(0)$ has vanishing projection on the null space of $\mc L$, i.e., $\mc P\psi(0)=\psi(0)$, the first term on the right hand side of \eqref{p5} is of order $\epsilon^{\frac 12}$ and exponentially small as $t\to\infty$. In particular, $u(T)$ is of order $\epsilon^{\frac 12}$ so that we can compute the displacement of the center in the time interval $[0,T]$ by solving \eqref{p5} in the linear approximation. We thus get
\begin{equation}
\label{p4.5}
\zeta_T = \zeta_0 - \gamma \int_0^1\!\mathrm{d} x\: \chi(x) u(x,T)+\cdots,
\end{equation} 
where $\gamma$ is defined in \eqref{gamma}.

\subsection{Computation of the variance and the drift}

The variance of the infinitesimal displacement $\zeta_T-\zeta_0$ can be effectively computed by considering the component of the noise $W^\epsilon$ in the direction spanned by the right eigenvector $\upbar{\rho}'$. Note indeed that the nonlinear term in \eqref{p5} is of order $\epsilon T$. Recalling \eqref{p5.5} and that $\chi \, \mathrm{e}^{t\mc L} = \chi$, 
\[
\begin{split} 
& \langle (\zeta_T -\zeta_0)^2\rangle \approx \epsilon \gamma^2 \Big\langle \Big(\int_0^1\!\mathrm{d} x\: \chi(x) W^\epsilon(x,T)\Big)^2\Big\rangle \approx \epsilon T \gamma^2 \int\!\mathrm{d} y\: \chi'(y) \Sigma(\upbar{\rho};y,y)\chi'(y)^\top \\ & \qquad = 2\epsilon T \gamma^2 \int\!\mathrm{d} y\: \big[\upbar{\rho}_A(1-\upbar{\rho}_A) (\upbar{\rho}_B')^2 + 2\upbar{\rho}_A\upbar{\rho}_B\upbar{\rho}_A'\upbar{\rho}_B' + \upbar{\rho}_B(1-\upbar{\rho}_B) (\upbar{\rho}_A')^2\big],
\end{split}
\]
where we evaluated \eqref{p5.5} with $\delta_\epsilon$ replaced by the true Dirac's $\delta$-function. By translation invariance, each step of the iterations has the same variance. Therefore, since the fluctuation of the infinitesimal displacements relative to different steps are also independent and mean zero, they sum up to a Brownian motion with variance $\sigma^2=\sigma^2(\beta; r_A,r_B)$ given by
\[
\sigma^2(\beta; r_A,r_B) = 2\gamma^2 \int\!\mathrm{d} y\: \big[\upbar{\rho}_A(1-\upbar{\rho}_A) (\upbar{\rho}_B')^2 + 2\upbar{\rho}_A\upbar{\rho}_B\upbar{\rho}_A'\upbar{\rho}_B' + \upbar{\rho}_B(1-\upbar{\rho}_B) (\upbar{\rho}_A')^2\big],
\]   
which has been computed in \cite[Eq.~(7)]{BD}.

As discussed in the Introduction, the non-linear term contributes to the phase fluctuations at times $\epsilon^{-1}$, by giving a constant drift to the resulting random motion. Our next aim is to identify this drift as a function of $\beta$ and of the total densities.

As for the variance, we discuss in detail the contribution picked up in the first step of the iteration. Indeed, again by translation invariance, each step of the iterations gives the same contribution to the drift. Since $u$ is of order $\epsilon^{\frac 12}$ and the non-linear term $\mc N_1$ in \eqref{p4.25} is homogenous of degree two, the latter gives a contribution of order $\epsilon T$ to the infinitesimal displacement $\zeta_T-\zeta_0$. Therefore, it sums up to a finite contribution at time-scale $\epsilon^{-1}$. 

By iterating once \eqref{p5} we get
\begin{equation*}
\begin{split}
u(T) & = \mathrm{e}^{T\mc L}\psi(0) + \sqrt\epsilon\, W^\epsilon(T) + \beta \int_0^T\!\mathrm{d} s\, \mathrm{e}^{(T-s)\mc L} \frac{\partial}{\partial x}\mc N_1(\mathrm{e}^{s\mc L}\psi(0)+ \sqrt\epsilon W^\epsilon(s)) + \cdots \\ & = \sqrt\epsilon\, W^\epsilon(T) + \beta \epsilon\int_0^T\!\mathrm{d} s\, \mathrm{e}^{(T-s)\mc L} \frac{\partial}{\partial x}\mc N_1(W^\epsilon(s)) + \cdots 
\end{split}
\end{equation*}
where we used again that the contribution from the initial condition is negligible for $T\gg 1$.  Plugging the last displayed into \eqref{p4.5} and taking expectation we obtain
\[
\begin{split}
\left\langle \zeta_T\right\rangle & \approx \zeta_0 -\beta\, \gamma\,\epsilon \int_0^T\!\mathrm{d} t \int_0^1\!\mathrm{d} x\:
\chi(x) \:\frac{\partial}{\partial x} \left\langle \mc N_1(W^\epsilon(x,t)) \right\rangle \\ 
& = \zeta_0 + \beta\, \gamma\,\epsilon
\int_0^T\!\mathrm{d} t \int_0^1\!\mathrm{d} x\: \chi'(x) \: \left\langle \mc N_1(W^\epsilon(x,t)) \right\rangle,
\end{split}
\]
where we used that $\chi$ is a left eigenvector of $\mc L$. Note also that the approximation in \eqref{p4} contributes to the displacement $\left\langle \zeta_T\right\rangle$ by a negligible amount. Replacing $\delta_\epsilon$ by the true Dirac's $\delta$-function in \eqref{p5.5} and recalling the form \eqref{p4.25} of the nonlinearity,
\[
\left\langle \zeta_T\right\rangle \approx \zeta_0+ \epsilon T \beta\,\gamma\, \frac 1T \int_0^T\!\mathrm{d} t\: \int_0^t\!\mathrm{d} s\:\int_0^1\!\mathrm{d} x\: \chi'(x) \begin{pmatrix} \mc K_{AA}(x,s) + 2 \mc K_{AB}(x,s) \\ -\mc K_{BB}(x,s) - 2 \mc K_{AB}(x,s) \end{pmatrix},
\]
where
\begin{equation}
\label{Z}
\mc K(x,t) = \int_0^1\!\mathrm{d} y\:\frac{\partial G}{\partial y}(x,y,t) \: \Sigma(\upbar{\rho};y,y)\: \frac{\partial G}{\partial y}(x,y,t)^\top.
\end{equation}
The drift $v=v(\beta;r_A,r_B)$ is thus given by 
\begin{equation}
\label{p7}
v(\beta;r_A,r_B) = \beta\gamma\int_0^\infty\!\mathrm{d} t\:\int_0^1\!\mathrm{d} x\: \chi'(x) \begin{pmatrix} \mc K_{AA}(x,t) + 2 \mc K_{AB}(x,t) \\ -\mc K_{BB}(x,t) - 2 \mc K_{AB}(x,t) \end{pmatrix}.
\end{equation}
This formula is our first main result; in the next sections we show that its right-hand side is finite and antisymmetric with respect to the exchange of the masses.

\section{Boundedness of the drift}
\label{sec:3}

The computations leading to the formula \eqref{p7} for the drift have been carried out informally, pretending the limits we took did exist. In fact, by looking at \eqref{p7}, it is not at all obvious that the time integral on its right hand side is finite. Without discussing in detail the limiting procedure performed before, in this section we show that such time integral is meaningful. There are two potential problems: the singularity of the kernel $\mc K$ for $t\downarrow 0$, which is the effect of the ultraviolet singularities mention before, and the convergence of the integral at infinity. 

We start with the analysis of the singularity around the origin. For $t$ small $G(x,y,t)$ behaves as the heat semigroup, i.e., 
\begin{equation*}
G(x,y,t) =  p_t(x-y) \begin{pmatrix} 1 & 0 \\ 0 & 1\end{pmatrix} + \mc H(x,y,t), 
\end{equation*}
where
\[
p_t(x-y) = \sum_{n\in\mb Z} \frac 1{\sqrt{4\pi t}} \, \mathrm{e}^{- \frac{(x-y+n)^2}{4t}}
\]
and $\mc H(x,y,t)$, $\frac{\partial\mc H}{\partial y}(x,y,t)$ are bounded as $t\downarrow 0$. Therefore, by \eqref{Z},  the integrand on the right hand side of \eqref{p7} should have the non integrable singularity $t^{-3/2}$  for $t$ close to $0$. As we next show, this divergence disappears due to a cancelation. We compute the (dangerous) contribution to the drift coming from the heat semigroup part of $G(x,y,t)$, i.e.,
\[
v^\mathrm{s} = \beta\gamma\int_0^\infty\!\mathrm{d} t\:\int_0^1\!\mathrm{d} x\: \chi'(x) \begin{pmatrix} \mc K_{AA}^\mathrm{s}(x,t) + 2 \mc K_{AB}^\mathrm{s}(x,t) \\ -\mc K_{BB}^\mathrm{s}(x,t) - 2 \mc K_{AB}^\mathrm{s}(x,t) \end{pmatrix}.
\]
with
\[
\mc K^\mathrm{s}(x,t) = \int_0^1\!\mathrm{d} y\:  \left[\frac{\partial p_t}{\partial y}(x-y)\right]^2\: \Sigma(\upbar{\rho};y,y) = \frac{1}{t^{3/2}} \int_0^1\!\mathrm{d} y\: g_t(x-y) \Sigma(\upbar{\rho};y,y) ,
\]
where the function
\[
g_t(z) =  \sum_{n,k\in\mb Z}  \frac{(z+n)(z+k)}{4\pi t^{3/2}}\,  \mathrm{e}^{- \frac{(z+k)^2+(z+n)^2}{4t}}
\]
behaves like $(4\sqrt\pi)^{-1}\delta(z)$ as $t\downarrow 0$. Therefore, 
\[
\mc K^\mathrm{s}(x,t) = \frac{1}{t^{3/2}} \left[\frac{1}{4\sqrt\pi}\Sigma(\upbar{\rho};x,x) + O(t)\right],
\]
so that
\[
\begin{split}
v^\mathrm{s} & = \beta\gamma\int_0^\infty\!\mathrm{d} t\:  \frac{1}{t^{3/2}}\int_0^1\!\mathrm{d} x\: \chi'(x) \: \left[ \frac{1}{4\sqrt\pi}\begin{pmatrix} \Sigma_{AA}(\upbar{\rho};x,x) + 2 \Sigma_{AB}(\upbar{\rho};x,x) \\  -\Sigma_{BB}(\upbar{\rho};x,x) - 2 \Sigma_{AB}(\upbar{\rho};x,x) \end{pmatrix} + O(t)\right] .
\end{split}
\]
On the other hand,
\[
\begin{split}
  & \int_0^1\!\mathrm{d} x\: \chi'(x) \: \begin{pmatrix} \Sigma_{AA}(\upbar{\rho};x,x) + 2 \Sigma_{AB}(\upbar{\rho};x,x) \\  -\Sigma_{BB}(\upbar{\rho};x,x) - 2 \Sigma_{AB}(\upbar{\rho};x,x) \end{pmatrix}  \\ & = \int_0^1\!\mathrm{d} x\:
  \upbar{\rho}_B'[2\upbar{\rho}_A(\upbar{\rho}_A-1) + 4 \upbar{\rho}_A\upbar{\rho}_B] +
  \upbar{\rho}_A'[2\upbar{\rho}_B(\upbar{\rho}_B-1) + 4 \upbar{\rho}_A\upbar{\rho}_B] \\ &
  = 2\int_0^1\!\mathrm{d}y \:
  (\upbar{\rho}_A^2\upbar{\rho}_B+\upbar{\rho}_A\upbar{\rho}_B^2 -\upbar{\rho}_A\upbar{\rho}_B
  )' = 0,
\end{split}
\]
i.e., the coefficient in front of the non integrable singularity $t^{-3/2}$ vanishes.

We now analyze the convergence of the integral \eqref{p7} at infinity. We notice that, by our assumptions on the spectrum of $\mc L$, 
\[
G(x,y,t) = \gamma\, \upbar{\rho}'(x)\,\chi(y) + \widehat G(x,y,t), 
\]
where the $2\times 2$ rank one matrix $\gamma\, \upbar{\rho}'(x)\,\chi(y)$
is the projection on the null space of $\mc L$, while the remainder
$\widehat G(x,y,t)$ is exponentially small as $t\to\infty$. Therefore,
plugging this decomposition in \eqref{Z} we get,
\begin{equation}
\label{p8}
\mc K(x,t)  = \kappa\, \upbar{\rho}'(x)\upbar{\rho}'(x)^\top + R(x,t),
\end{equation}
where $\kappa = \gamma^2\int_0^1\!\mathrm{d} y\:\chi'(y)\Sigma(\upbar{\rho};y,y)\chi'(y)^\top$, while the remainder $R(x,t)$ is integrable as $t\to\infty$.  

By \eqref{p4.25}, inserting \eqref{p8} in \eqref{p7} and integrating by parts, we deduce that the drift $v$ is finite provided that  
\begin{equation}
\label{p9bis}
\int_0^1\!\mathrm{d} x\: \chi' \begin{pmatrix} (\upbar{\rho}_A')^2+2\upbar{\rho}_A'\upbar{\rho}_B'
\\ -(\upbar{\rho}_B')^2-2\upbar{\rho}_A'\upbar{\rho}_B'\end{pmatrix} = 0. 
\end{equation}
As $\chi=(\upbar{\rho}_B,-\upbar{\rho}_A)$, the above conditions reads, 
\begin{equation}
\label{p9}
3 \int_0^1\!\mathrm{d} x\: \big[ (\upbar{\rho}_A')^2\upbar{\rho}_B' +
\upbar{\rho}_A'(\upbar{\rho}_B')^2 \big] = 0.
\end{equation}
To prove \eqref{p9}, we denote by $I$ the integral on the left-hand side. Recalling that $\rho_C=1-\rho_A-\rho_B$ and integrating by parts, we then get
 \begin{equation}
 \label{f}
 \begin{split}
 I & = - 3\int_0^1\!\mathrm{d} x\: \upbar{\rho}_A'\upbar{\rho}_B'\upbar{\rho}_C' = 
 \int_0^1\!\mathrm{d} x\: \big[\upbar{\rho}_A(\upbar{\rho}_B'\upbar{\rho}_C')' + \upbar{\rho}_B(\upbar{\rho}_A'\upbar{\rho}_C')' + \upbar{\rho}_C(\upbar{\rho}_A'\upbar{\rho}_B')' \big]  \\ & = \int_0^1\!\mathrm{d} x\: \big[(\upbar{\rho}_A\upbar{\rho}_B)'\upbar{\rho}_C'' + (\upbar{\rho}_A\upbar{\rho}_C)'\upbar{\rho}_B''  +(\upbar{\rho}_B\upbar{\rho}_C)'\upbar{\rho}_A''\big].
\end{split}
 \end{equation}
On the other hand, \eqref{ss} in term of the triple $(\upbar{\rho}_A,\upbar{\rho}_B,\upbar{\rho}_C)$ reads,
\[
\begin{cases} \upbar{\rho}_A'' = -\beta[(\upbar{\rho}_A\upbar{\rho}_B)'-(\upbar{\rho}_A\upbar{\rho}_C)'], \\ \upbar{\rho}_B'' = -\beta[(\upbar{\rho}_B\upbar{\rho}_C)'-(\upbar{\rho}_B\upbar{\rho}_A)'] ,\\ \upbar{\rho}_C'' = -\beta[(\upbar{\rho}_C\upbar{\rho}_A)'-(\upbar{\rho}_C\upbar{\rho}_B)']. \end{cases}
\]
Substituting the  above relations in \eqref{f} the identity $I=0$ follows.

\section{Antisymmetry of the drift}
\label{s:5}

In this section we show that, as thought by Bodineau and Derrida \cite[\S~6]{BD}, the drift  
is antisymmetric with respect to the exchange of the total densities. More precisely, with a slight abuse of notation, let $v=v(\beta,r_A,r_B,r_C)$ be the drift in \eqref{p7}. Then  $v(\beta,r_{\tau A}, r_{\tau B}, r_{\tau C}) = - v(\beta,r_A,r_B,r_C)$ where $\tau$ is a transposition acting on $\{A,B,C\}$.  

As a matter of fact, we present two independent arguments. The first relies on a direct analysis on the expression \eqref{p7}, while in the second we exploit a symmetry of the underlying microscopic  dynamics. 

\subsection{Macroscopic computation}

On the set of profiles $\rho = \begin{pmatrix} \rho_A \\ \rho_B \end{pmatrix}$ we introduce
the involution $\Theta$ defined by  
\[
(\Theta \rho)\, (x) = \begin{pmatrix} \rho_B(-x) \\ \rho_A(-x) \end{pmatrix}.
\]
In the language of high energy physics one may interpret $\Theta$ as a CP symmetry. 

We then observe that if we include the dependence on the total
densities $r_A,r_B$ in the notation, the symmetry
\[
\Theta \upbar{\rho}(r_A,r_B) = \upbar{\rho}(r_B,r_A)
\]
holds modulo a translation on the torus (look at \eqref{ss}).

Denoting by $\widetilde{\mc L}$ the linear operator defined as $\mc L$
but with the matrix function $\mc B$ replaced by 
\begin{equation}
\label{p3bis}
\Theta \mc B =\begin{pmatrix} 2(\Theta\upbar{\rho})_A+2(\Theta\upbar{\rho})_B-1 &
2(\Theta\upbar{\rho})_A \\ -2(\Theta\upbar{\rho})_B &
-2(\Theta\upbar{\rho})_A-2(\Theta\upbar{\rho})_B+1 \end{pmatrix}. 
\end{equation}
it is easy to see that $\Theta\circ\mc L = \widetilde{\mc L} \circ
\Theta$. This implies that if $\psi(x,t)$ is solution to
$\partial_t\psi = \mc L \psi$ then $\widetilde{\psi}(x,t) =
(\Theta\psi(\cdot,t))\,(x)$ is solution to $\partial_t
\widetilde{\psi} = \widetilde{\mc L} \widetilde{\psi}$. In particular, 
\[
G(x,y,t) = \begin{pmatrix} 0 & 1 \\ 1 & 0 \end{pmatrix} \widetilde{G}(-x,-y,t) \begin{pmatrix} 0 & 1 \\
1 & 0 \end{pmatrix} , 
\]
where $\widetilde{G}(x,y,t)=\mathrm{e}^{t\widetilde{\mc L}}(x,y)$ denotes
the fundamental solution associated to $\widetilde{\mc L}$. On the
other hand, recalling \eqref{p1.5},  
\[
\widetilde{\Sigma}(\upbar{\rho};y,y) = \Sigma(\Theta\upbar{\rho};y,y) = \begin{pmatrix} 0
& 1 \\ 1 & 0 \end{pmatrix} \Sigma(\upbar{\rho};-y,-y) \begin{pmatrix} 0 & 1 \\ 1 & 0 \end{pmatrix}, 
\]
whence, by \eqref{Z},
\[
\mc K(x,t) = \begin{pmatrix} 0 & 1 \\ 1 & 0 \end{pmatrix} \widetilde{\mc K}(-x,t)\begin{pmatrix} 0 & 1 \\ 1 & 0 \end{pmatrix} 
\]
where 
\[
\widetilde{\mc K}(x,t) = \int_0^1\!\mathrm{d} y\:\frac{\partial \widetilde{G}}{\partial y}(x,y,t)  \:\widetilde{\Sigma}(\upbar{\rho};y,y)\: \frac{\partial \widetilde{G}}{\partial y}(x,y,t)^\top.
\]
Therefore, by \eqref{p7}
\begin{equation}
\label{p7bis}
\begin{split}
v(\beta;r_A,r_B) & = \beta\gamma\int_0^\infty\!\mathrm{d} t\:\int_0^1\!\mathrm{d} x\: \chi'(x) \begin{pmatrix} \widetilde{\mc K}_{BB}(-x,t) + 2 \widetilde{\mc K}_{AB}(-x,t) \\ -\widetilde{\mc K}_{AA}(-x,t) - 2 \widetilde{\mc K}_{AB}(-x,t) \end{pmatrix}\\ & = - \beta\gamma\int_0^\infty\!\mathrm{d} t\:\int_0^1\!\mathrm{d} x\: (\Theta\chi')\,(x) \:  \begin{pmatrix} \widetilde{\mc K}_{AA}(x,t) + 2 \widetilde{\mc K}_{AB}(x,t) \\ -\widetilde{\mc K}_{BB}(x,t) - 2 \widetilde{\mc K}_{AB}(x,t) \end{pmatrix} \\ & = - \beta\gamma\int_0^\infty\!\mathrm{d} t\:\int_0^1\!\mathrm{d} x\: \widetilde{\chi}'(x)  \:  \begin{pmatrix} \widetilde{\mc K}_{AA}(x,t) + 2 \widetilde{\mc K}_{AB}(x,t) \\ -\widetilde{\mc K}_{BB}(x,t) - 2 \widetilde{\mc K}_{AB}(x,t) \end{pmatrix},
\end{split}
\end{equation}
where $\widetilde\chi = (\Theta\upbar{\rho}_B,-\Theta\upbar{\rho}_A)$ is the
left eigenvector of $\widetilde{\mc L}$ with zero eigenvalue and the
identity $\widetilde{\chi}' = \Theta\chi'$ is
straightforward. Recalling $\Theta \upbar{\rho}(r_A,r_B) =
\upbar{\rho}(r_B,r_A)$ holds modulo a translation on the torus, by
\eqref{p7bis} we conclude $v(\beta;r_A,r_B) = -v(\beta;r_B,r_A)$. 
Note indeed that by arguing as above (it is in fact easier) it holds 
$\gamma(r_A,r_B)= \gamma(r_B,r_A)$. By the arbitrariness of the labels' choice this shows the stated antisymmetry property.

\subsection{A microscopic symmetry}

From a mathematical point of view, the ABC model is a continuous time Markov chain on the state space $\Omega=\{A,B,C\}^{\mb Z_L}$, where $\mb Z_L=\{0,\ldots,L-1\}$ is the ring of the integers modulo $L$. Given $\zeta\in\Omega$, the species at site $i$ is therefore $\zeta(i)\in \{A,B,C\}$. For $\beta\in\mb R$ the dynamical rules are specified by the generator $\mb{L}^\beta$ that acts on observables $f\colon \Omega\to \mb{R}$ as
\begin{equation}
\label{gen}
  \mb{L}^\beta f \, (\zeta)  = \sum_{i\in\mb Z_L} c^\beta_i(\zeta) 
  \big[ f(\zeta^{i,i+1}) - f(\zeta)\big],
\end{equation}
where $\zeta^{i,i+1}$ is the configuration obtained from $\zeta$ by exchanging the species at sites $i$ and $i+1$ and the jump rates $c^\beta_i$ are given by 
\begin{equation}
\label{rate}
  c^\beta_i(\zeta) =
  \begin{cases}
    \exp\{ \frac \beta{2L}\}
    & \text{ if $\; (\zeta(i),\zeta(i+1)) \in 
      \{ (A,C), (C,B), (B,A)\}$,} \\  
    \exp\{ -\frac \beta{2L}\big\} 
    &\text{ if $\; (\zeta(i),\zeta(i+1)) \in 
      \{ (A,B), (B,C), (C,A)\}$.} 
  \end{cases}
\end{equation}

We next prove a symmetry property of the microscopic stochastic dynamics with respect to a suitable involution defined on the state space $\Omega$. The antisymmetry of the macroscopic drift will be then shown to be a consequence of such symmetry. Let $\tau$ be a transposition of the species' labels and denote by $\tau\alpha\in \{A,B,C\}$ the image of $\alpha\in\{A,B,C\}$. We associate to each $\tau$ the involution $\Theta_\tau\colon\Omega\to\Omega$ defined by
\[
\Theta_\tau\zeta\,(i) = \tau\zeta\, (-i),
\]
which induces a natural transformation on observables $f\colon \Omega\to \mb{R}$ by setting $\Theta_\tau f(\zeta) = f(\Theta_\tau\zeta)$. We claim that the microscopic dynamics satisfies, 
\begin{equation}
\label{sim}
\Theta_\tau \circ \mb{L}^{-\beta}  =  \mb{L}^\beta \circ \Theta_\tau.
\end{equation}
To prove the relationship \eqref{sim}, observe that $(\alpha,\alpha')\in \{ (A,C), (C,B), (B,A)\}$ if and only if  $(\tau\alpha,\tau\alpha') \in \{ (A,B), (B,C), (C,A)\}$. Whence, by \eqref{rate}, 
\[
c^{-\beta}_i(\Theta_\tau\zeta) = c^\beta_{-(i+1)}(\zeta).
\]
Recalling \eqref{gen} we deduce,
\[
\Theta_\tau \circ \mb{L}^{-\beta} \, f\, (\zeta)  = \sum_{i\in\mb Z_L} c^\beta_{-(i+1)}(\zeta) 
\big[ f((\Theta_\tau\zeta)^{i,i+1}) - f(\Theta_\tau\zeta)\big].
\]
Therefore, as $(\Theta_\tau\zeta)^{i,i+1} = \Theta_\tau\zeta^{-(i+1),-i}$, 
\[
\begin{split}
\Theta_\tau \circ \mb{L}^{-\beta} \, f \, (\zeta) &  = \sum_{i\in\mb Z_L} c^\beta_{-(i+1)}(\zeta) 
\big[ f(\Theta_\tau\zeta^{-(i+1),-i}) - f(\Theta_\tau\zeta)\big] \\ & =  \sum_{j\in\mb Z_L} c^\beta_j(\zeta) \big[ f(\Theta_\tau\zeta^{j,j+1}) - f(\Theta_\tau\zeta)\big]  \\ & = \mb{L}^\beta \circ \Theta_\tau \, f \, (\zeta).
\end{split}
\]

The relationship \eqref{sim} implies the corresponding symmetry on the statistics of the paths of the ABC model. In particular, if the process has a macroscopic drift $v = v(\beta;r_A,r_B,r_C)$ then it necessarily  satisfies $v(\beta;r_A,r_B,r_C) = v(-\beta;r_{\tau A},r_{\tau B},r_{\tau C})$. The antisymmetry of the drift with respect to the transposition of the species' labels now holds provided $v$ is an odd function of $\beta$. Recalling that $v$ is given by the expression \eqref{p7}, this can be easily verified.

\section{Critical behavior}
\label{s:cb}

Consider values of the masses $r=(r_A,r_B,r_C)$ that satisfy the condition $r^2>\frac 12$ and $r^2<2(r_A^3+r_B^3+r_C^3)$. As discussed in \cite{8,CM}, in this case the transition at $\beta_r = 2\pi/\sqrt{1-2r^2}$ is of second order. In particular, increasing $\beta$ above the threshold $\beta_r$, the homogeneous profile $(r_A,r_B,r_C)$ looses its linear stability and bifurcates giving rise to the stationary solutions $\upbar\rho_z$, $z\in \bf T$. In this section we analyze the behavior of the drift $v(\beta; r_A,r_B)$ in \eqref{p7} as $\beta\downarrow \beta_r$. The corresponding analysis for the variance $\sigma^2=\sigma^2(\beta; r_A,r_B)$ has been carried out in \cite{BD}, yielding $\sigma^2\approx (\beta-\beta_r)^{-1}$. 

We start by the expansion of the stationary profiles $\upbar\rho_z$ as $\beta\downarrow \beta_r$. The first order correction is computed in \cite{8}, however we shall need also the second order. The details of this computation are given in Appendix \ref{app:c1}. 

Analogously to the notation in \cite{8}, we introduce the real parameters,
\begin{equation}
\label{tp} 
\theta = \frac{\beta-\beta_r}{\beta_r}, \qquad \psi(\theta) = \frac{1-2r^2}{\sqrt{2r^2-4(r_A^3+r_B^3+r_C^3)}}\,\theta^{1/2}.
\end{equation}
By defining 
\begin{equation}
\label{phi}
\phi = \mathrm{arg}(1-2r_A-2r_B-\mathrm{i}\sqrt{1-2r^2}),
\end{equation}
the expansion of the stationary solution is
\begin{equation}
\label{r=}
\begin{split}
\upbar\rho_z(x) & = \begin{pmatrix} r_A \\ r_B \end{pmatrix} + \psi(\theta) \begin{pmatrix} \sqrt{r_A} \\ \sqrt{r_B} \mathrm{e}^{\mathrm{i}\phi} \end{pmatrix} \mathrm{e}^{2\pi\mathrm{i} (x-z)} \\ & \quad + \frac{\psi(\theta)^2}{1-2r^2} \begin{pmatrix} r_A(1-2r_A) \\ r_B(1-2r_B) \mathrm{e}^{2\mathrm{i}\phi} \end{pmatrix}\mathrm{e}^{4\pi\mathrm{i} (x-z)} + \mathrm{c.c.} + O(\theta^{3/2}).
\end{split}
\end{equation}
In particular, $\phi/(2\pi)$ represents the phase shift between the $A$ and $B$ species.

To compute the kernel of $\mathrm{e}^{t\mc L_z}$ we regard the operator $\mc L_z$ as a perturbation of the differential operator with constant coefficients that is obtained by linearizing \eqref{hy} around the homogenous profile $r$. The corresponding perturbation theory is detailed in Appendix \ref{app:c2} and yields the following. It is convenient to introduce the vectors,
\begin{equation}
\label{LamIup}
\Lambda = \sqrt{\frac{r_A}{1-2r^2}} \begin{pmatrix} \sqrt{r_A} \\ \sqrt{r_B} \, \mathrm{e}^{\mathrm{i}\phi} \end{pmatrix},\quad \Upsilon = \frac{2}{1-2r^2}  \sqrt{\frac{r_A}{1-2r^2}} \begin{pmatrix} r_A(1-2r_A) \\ r_B (1-2r_B) \; \mathrm{e}^{2\mathrm{i}\phi} \end{pmatrix}.
\end{equation}
As already discussed, $0$ is a simple eigenvalues of $\mc L_z$ and the corresponding right and left eigenvectors are given by $\hat e_z(x) = \hat e(x-z)$ and $\hat w_z(x) = \hat w(x-z)$ where
\[
\begin{split}
\hat e(x) & = \frac{1}{2\sqrt{r_A}} \left(\mathrm{i}\Lambda\,\mathrm{e}^{2\pi\mathrm{i} x} + \psi(\theta)\, \mathrm{i} \Upsilon\, \mathrm{e}^{4\pi\mathrm{i} x} \right) + \mathrm{c.c.} + O(\theta),\\ \hat w(x) & = \sqrt{\frac{1-2r^2}{r_A}} \left[ - \begin{pmatrix} 0 & 1 \\ -1 & 0 \end{pmatrix} \Lambda \,\mathrm{e}^{2\pi\mathrm{i} x} + \frac 12 \psi(\theta) \begin{pmatrix} 0 & 1 \\ -1 & 0 \end{pmatrix}\Upsilon \, \mathrm{e}^{4\pi\mathrm{i} x} \right]^\top+ \mathrm{c.c.}+ O(\theta).
\end{split}
\]
Moreover, $\mc L_z$ has a small negative simple eigenvalue $\lambda=-8\pi^2\theta+O(\theta^{3/2})$; the corresponding right and left eigenvectors are given by $e_z(x) = e(x-z)$ and $w_z(x) = w(x-z)$ where
\[
\begin{split}
e(x) & = \Lambda\, \mathrm{e}^{2\pi\mathrm{i} x} + \psi(\theta) \Upsilon\, \mathrm{e}^{4\pi\mathrm{i} x} + \mathrm{c.c.}+ O(\theta), \\ w(x) & = \frac{\sqrt{1-2r^2}}{2r_A} \left[ \begin{pmatrix} 0 & 1 \\ -1 & 0 \end{pmatrix} \mathrm{i} \Lambda \,\mathrm{e}^{2\pi\mathrm{i} x} + \frac 12 \psi(\theta) \begin{pmatrix} 0 & 1 \\ -1 & 0 \end{pmatrix}\mathrm{i} \Upsilon \, \mathrm{e}^{4\pi\mathrm{i} x} \right]^\top + \mathrm{c.c.}+ O(\theta).
\end{split}
\]
Finally, the other eigenvalues of $\mc L_z$ have negative real part, bounded away from zero  uniformly in $\theta$. As follows from the discussion in Section \ref{s:1}, the right and left eigenfunctions corresponding to the zero eigenvalue are given by $\upbar\rho'(x)$ and $\chi(x)$, see \eqref{p35}. Indeed, $\hat e = \psi\gamma\upbar\rho'$ and $\hat w = \psi^{-1} \chi$. Observe also that the above eigenvectors are bi-orthonormal. Namely, by introducing the canonical pairing $\lan u|v\ran = \int\!\mathrm{d} y\: u(y)v(y)$ we have,
\[
\lan w|e \ran = \lan \hat w| \hat e \ran = 1, \qquad \lan w| \hat e \ran = \lan \hat w| e \ran = 0.
\]

We now proceed to the expansion of the drift $v(\beta;r_A,r_B)$ in \eqref{p7}. In view of the previous expansion, 
\begin{equation}
\label{GE}
\frac{\partial G}{\partial y}(x,y,t) = \hat e(x)\hat w'(y) + \mathrm{e}^{\lambda t} e(x) w'(y) + R(x,y,t),
\end{equation}
with
\[
R(x,y,t) = \sum_{n>1} \mathrm{e}^{\lambda_n t} e_n(x) w_n'(y),
\]
where $\{\lambda_n\}_{n>1}$ is the rest of the spectrum of $\mc L$ and $e_n,w_n$ are the corresponding eigenvectors. In particular, $R(x,y,t)=R_0(x,y,t) + O(\theta^{1/2}\mathrm{e}^{-ct})$, where $R_0$ is defined by computing the above displayed expression for $\theta=0$ and $c>0$ is independent of $\theta$. Observe $R_0$ does not contain the Fourier modes $\mathrm{e}^{\pm 2\pi \mathrm{i}x}$. Note that we did not expand with respect to $\theta$ the first term in the right-hand side of \eqref{GE} as we already know it does not contribute to the drift, see Section \ref{sec:3}. We next observe that 
\[
\Sigma(\upbar\rho;x,x)  = \Sigma_0 + \psi(\theta) \Sigma_1 \mathrm{e}^{2\pi\mathrm{i} x} + O(\theta) + \mathrm{c.c.}, \quad \Sigma_0 = \begin{pmatrix} 2r_A(1-r_A) & - 2r_Ar_B \\ -2r_Ar_B & 2r_B(1-r_B) \end{pmatrix},
\]
for some matrix $\Sigma_1$ with constant entries. Inserting the expansions in \eqref{Z}, we get
\[
\mc K(x,t) \! = \! K_0 \hat e(x) \hat e(x)^\top + K_1 \mathrm{e}^{\lambda t}\left[e(x)\hat e(x)^\top + \hat e(x) e(x)^\top\right] + K_2 \,\mathrm{e}^{2\lambda t} e(x)e(x)^\top + \mc R(x,t),
\]
where $K_0 = \lan \hat w'| \Sigma(\bar\rho)(\hat w')^T \ran$, and, by a computations, $K_1 = \lan \hat w'| \Sigma(\bar\rho)(w')^T \ran  = O(\theta)$, $K_2= 12\pi^2r_Br_C+O(\theta)$. Finally, $\mc R(x,t) = \mc R_0(x,t) + O(\theta^{1/2}\mathrm{e}^{-ct})$, where $\mc R_0$ does not contain the Fourier modes $\mathrm{e}^{\pm 2\pi \mathrm{i}x}$.

Therefore, by \eqref{p7},
\begin{equation*}
\begin{split}
v(\beta;r_A,r_B) & = \beta\gamma\int_0^\infty\!\mathrm{d} t\: \left\langle \chi'\,\bigg|\begin{pmatrix} \mc K_{AA}(\cdot,t) + 2 \mc K_{AB}(\cdot,t) \\ -\mc K_{BB}(\cdot,t) - 2 \mc K_{AB}(\cdot,t) \end{pmatrix}\right\rangle \\ & = \beta\gamma\left[ - K_1\frac{U_1}{\lambda} - K_2\frac{U_2}{2\lambda} + O(\theta) \right],
\end{split}
\end{equation*}
where, letting $H(x) = e(x)\hat e(x)^\top + \hat e(x) e(x)^\top$, 
\[
U_1 = \left\langle \chi'\,\bigg| \begin{pmatrix} 2H_{AA} + 2 H_{AB} + 2H_{BA}  \\ - 2H_{BB}- 2 H_{AB}- 2H_{BA}\end{pmatrix} \right\rangle, \qquad U_2 = \left\langle \chi'\,\bigg|\begin{pmatrix} e_Ae_A + 2 e_Ae_B  \\ - e_Be_B- 2 e_Ae_B\end{pmatrix}\right\rangle.
\]
In view of the expansions of the right eigenvalues $\hat e(x), e(x)$,
\[
H(x) = \frac{\mathrm{i}}{\sqrt{r_A}} \Lambda\Lambda^\top \,\mathrm{e}^{4\pi\mathrm{i} x} + \mathrm{c.c.} + O(\theta),
\]
whence $U_1=O(\theta)$. Moreover, 
\[
e(x)e(x)^\top =\Lambda\Lambda^\top \,\mathrm{e}^{4\pi\mathrm{i} x} + \psi(\theta) (\Upsilon\bar\Lambda^\top + \bar\Lambda\Upsilon^\top) \,\mathrm{e}^{2\pi\mathrm{i} x}  + \mathrm{c.c.} + \cdots
\]
where the remainder does not involve modes $\mathrm{e}^{\pm 4\pi\mathrm{i} x}$ up to order $\theta^{3/2}$ (this follows from the analysis in Appendix \ref{app:c2}). Whence, by an explicit computation that exploits a cancelation due to the fact that $\Upsilon_A\bar\Lambda_B^2 + 2 \Upsilon_A\bar \Lambda_A\bar\Lambda_B + 2 \Upsilon_B\bar \Lambda_A\bar\Lambda_B + \Upsilon_B\bar \Lambda_A^2$ is real, we obtain $U_2=O(\theta^2)$. 

Recalling \eqref{gamma}, by \eqref{r=} we get $\gamma^{-1} = -4\pi\psi(\theta)^2 \sqrt{1-2r^2}
+ O(\theta^{3/2})$. Since $\lambda = -8\pi^2\theta + O(\theta^{3/2})$ we finally deduce that the drift $v(\beta;r_A,r_B)$ does not diverge as $\beta\downarrow\beta_r$. Since the variance $\sigma^2(\beta;r_A,r_B)$ diverges as $(\beta-\beta_r)^{-1}$, we conclude that the drift of phase fluctuations is not relevant in the critical regime. 

\appendix

\normalsize

\section{Nonexistence of traveling waves}
\label{s:ntw}
     
In this appendix we prove that the system \eqref{hy} does not admit traveling waves. Considering the case relevant for the ABC dynamics in which $\rho_A + \rho_B+\rho_C=1$, a  traveling wave of speed $c$ is a solution of \eqref{hy} of the form,
\[
\rho(x,t) = \begin{pmatrix} Y_A(x-ct) \\ Y_B(x-ct) \\ 1-Y_A(x-ct)-Y_B(x-ct)\end{pmatrix},
\]
where, recalling \eqref{N}, $Y = \begin{pmatrix} Y_A \\ Y_B \end{pmatrix}$ is a noncostant one-periodic solution to
\begin{equation}
\label{Y}
-c\, Y' = Y'' +\beta \, \mc N(Y)'.
\end{equation}
By integration, it follows that $Y$ is a noncostant one-periodic solution of the following first order planar system,
\begin{equation}
\label{ps}
Y' =  F_K(Y),
\end{equation}
where $F_K(Y) = K-cY-\beta\mc N(Y)$ for some constant vector $K=\begin{pmatrix} K_A \\ K_B \end{pmatrix}$. 

We claim that for $c\ne 0$ the system \eqref{ps} does not admit periodic noncostant solutions. In particular, there are not one-periodic solution, thus proving the absence of traveling waves for the original system \eqref{hy}. 
The key observation is that the divergence of the vector field $F_K$ has a definite sign for $c\ne 0$, more precisely $\mathrm{div} F_K = -2c$. This turns out to be an obstruction to the existence of noncostant periodic solution to system \eqref{ps}, as shown by the following argument (known in literature as Bendixson's criterium, see, e.g., \cite[\S~II.4.67]{NS}). Indeed, assume by contradiction that such periodic solution exists and denote by $\Lambda$ the bounded region of the plane  delimited by its orbit. Let $\hat n$ be the outer normal to $\Lambda$. By the divergence theorem,
\[
\oint_{\partial\Lambda}F_K\cdot\hat n = \int_\Lambda\!\mathrm{div} F_K = 2c|\Lambda|,
\]
where $|\Lambda|$ is the area of $\Lambda$. Since the boundary $\partial\Lambda$ is an orbit of \eqref{ps}, the vector field $F_K$ is tangential to it, i.e., $F_K\perp \hat n$ on $\partial\Lambda$. This implies $\oint_{\partial\Lambda}F_K\cdot\hat n = 0$, which yields the desired contradiction for $c\ne 0$. 

\section{How to guess fluctuating hydrodynamics}
\label{s:md}

We here briefly discuss, without even attempting any mathematical justification, how the fluctuating hydrodynamics \eqref{fhy}--\eqref{p1.5-1} can be inferred from the underlying
microscopic Markovian dynamics. We just outline the basic computation involved, referring 
to \cite{S} for the general principles of fluctuating hydrodynamics. A somewhat alternative argument to the one detailed below is presented in \cite{BD,bdlw,8}. 

Recalling the notation for the ABC dynamics introduced in Section \ref{s:5}, 
the occupation numbers of the species are given by
$\sigma_\alpha(i)=\id_{\alpha} (\zeta(i))$, $\alpha\in\{A,B,C\}$,
$i\in \mb Z_L$. From these variables we construct the empirical
densities of the species as
\begin{equation*}
  \pi_\alpha = \frac 1L \sum_{i\in \mb Z_L}\sigma_\alpha(i)
  \delta_{i/L}, \qquad \alpha\in\{A,B,C\},
\end{equation*}
that we regard as a random measure on the unit torus $\mb T$.

To infer the evolution of the empirical densities, we recall that if
$f\colon \Omega\to \mb R$ is an observable then, as follows from the
theory of Markov chains, its expected infinitesimal increment
is $\mb{L}^\beta f$, i.e., conditionally on $\zeta(t)$ it holds 
\begin{equation*}
  \big\langle f(\zeta(t+\mathrm{d}t)) - f(\zeta(t)) \big\rangle
  = \mb{L}^\beta f (\zeta(t))\mathrm{d}t + o(\mathrm{d}t).
\end{equation*}
In order to perform this computation for the empirical densities it is
convenient to introduce smooth test functions $J_\alpha\colon \mb T
\to \mb R$. By setting
\begin{equation*}
  \langle \pi_\alpha, J_\alpha\rangle =
  \frac 1L \sum_{i\in \mb Z_L}J_\alpha\big( \tfrac iL\big)  \sigma_\alpha(i),
\end{equation*}
simple computations then show,
\begin{equation*}
  \begin{split}
 \mb{L}^\beta \langle \pi_\alpha, J_\alpha\rangle
&  = \frac 1L \sum_{i\in \mb Z_L}
  \big[ J_\alpha\big( \tfrac {i+1}L\big)- J_\alpha\big( \tfrac iL\big)\big]   
  \big\{ \sigma_\alpha(i) 
  [ \mathrm{e}^{-\frac \beta{2L}} \sigma_{\alpha+1}(i+1) + \mathrm{e}^{\frac \beta{2L}}\sigma_{\alpha+2}(i+1)]
  \\
  &\quad  - \, [\mathrm{e}^{\frac \beta{2L}}\sigma_{\alpha+1}(i) + \mathrm{e}^{-\frac \beta{2L}}  \sigma_{\alpha+2}(i)]
  \sigma_\alpha(i+1) \big\},
  \end{split}
\end{equation*}
where the summation in the species labels $\alpha$ is modulo three. 
By approximating discrete gradients with continuous derivatives,
expanding the exponential, and summing by parts, we then obtain
\begin{equation*}
  \mb{L}^\beta \langle \pi_\alpha, J_\alpha\rangle
  \approx 
  \frac 1{L^2} \frac 1L \sum_{i\in \mb Z_L}
  \big\{ J_\alpha''\big( \tfrac iL\big) \sigma_{\alpha+1}(i) 
  -\beta \, J_\alpha'\big( \tfrac iL\big)
  \sigma_\alpha(i) \big[ \sigma_{\alpha+1}(i+1)- \sigma_{\alpha+2}(i+1)\big]
  \big\}.
\end{equation*}
Provided that we could replace the local product
$\sigma_\alpha(i)\sigma_{\alpha'}(i+1)$ above with the corresponding 
product of the local densities, we then identify the drift term in the
fluctuating hydrodynamic equation \eqref{fhy}. 
While this factorization hypotheses has been widely used, see e.g.,
\cite{bdlw}, we mention that it could be rigorously justified by the
methods of the hydrodynamical limits. More precisely, if we
consider an initial configuration of the species that is associated to
some density profile
$\rho(0)=(\rho_A(0),\rho_B(0),\rho_C(0))$
then the empirical densities at times $O(L^2)$ converges as
$L\to\infty$ to the solution of the hydrodynamic equation \eqref{hy}
with initial datum $\rho(0)$.

We emphasize that in order to justify the use of the fluctuating
hydrodynamics for the computation of the drift of phase fluctuations, we
would need the validity of the factorization assumption up to times
$O(L^3)$, certainly a most challenging issue in the context of
hydrodynamical limits.

To identify the fluctuations of the empirical densities, we recall
that, according to the general theory of Markov chains, if $f,g \colon
\Omega\to \mb R$ are two observables then, conditionally on $\zeta(t)$,
\begin{equation*}
  \begin{split}
  &\Big\langle 
  \big[ f(\zeta(t+\mathrm{d}t))-f(\zeta(t)) -\mb{L}^\beta f(\zeta(t)) \mathrm{d}t  \big]
  \cdot \big[ g(\zeta(t+\mathrm{d}t))-g(\zeta(t)) -\mb{L}^\beta g(\zeta(t)) \mathrm{d}t  \big]
  \Big\rangle
  \\
  &\qquad
  =\big[ \mb{L}^\beta f g \,(\zeta(t)) - f(\zeta(t))  \,\mb{L}^\beta g(\zeta(t))
   -g(\zeta(t))  \, \mb{L}^\beta f (\zeta(t)) \big] \, \mathrm{d}t + o(\mathrm{d}t).
  \end{split}
\end{equation*}
The fluctuations of the empirical densities $\pi_\alpha$ are therefore
characterized by the quadratic form,
\begin{equation*}
  \begin{split}
  & \Gamma_{\alpha,\alpha'}(J_\alpha,J_{\alpha'})=
  \mb{L}^\beta \big[ \langle\pi_\alpha, J_\alpha\rangle 
  \langle\pi_{\alpha'}, J_{\alpha'}\rangle \big]
  \\
  &\qquad  -  \langle\pi_\alpha, J_\alpha\rangle 
  \mb{L}^\beta \langle\pi_{\alpha'}, J_{\alpha'}\rangle  
  - \langle\pi_{\alpha'}, J_{\alpha'}\rangle 
  \mb{L}^\beta \langle\pi_{\alpha}, J_{\alpha}\rangle  
  \end{split}
  \quad \alpha,\alpha'\in\{A,B,C\},
\end{equation*}
where as before $J_\alpha:\mb T\to \mb R$ are smooth test functions. 
By straightforward computations, for $\alpha'=\alpha$,
\begin{equation*}
  \begin{split}
  &\Gamma_{\alpha,\alpha}(J_\alpha,J_{\alpha}) 
  =  \frac 1{L^2} \sum_{i\in \mb Z_L}
  \big[ J_\alpha\big( \tfrac {i+1}L\big)- J_\alpha\big( \tfrac iL\big)\big]^2   
  \big\{ \sigma_\alpha(i)   \big[ \mathrm{e}^{-\frac \beta{2L}} \sigma_{\alpha+1}(i+1) 
   \\
  &\qquad + \mathrm{e}^{\frac \beta{2L}}\sigma_{\alpha+2}(i+1)\big]
  +\big[\mathrm{e}^{\frac \beta{2L}}\sigma_{\alpha+1}(i) + \mathrm{e}^{-\frac \beta{2L}}\sigma_{\alpha+2}(i)\big]
  \sigma_\alpha(i+1) \big\}
  \\ 
  &\; 
  \approx
  \frac 1{L^2} \frac 1{L^2} \sum_{i\in \mb Z_L}J_\alpha'\big( \tfrac iL\big)^2
  \\
  &\quad \times 
  \big\{ \sigma_\alpha(i) 
  \big[ \sigma_{\alpha+1}(i+1) + \sigma_{\alpha+2}(i+1)\big]
  +\big[\sigma_{\alpha+1}(i) + \sigma_{\alpha+2}(i)\big]
  \sigma_\alpha(i+1) \big\},
  \end{split}
\end{equation*}
while, for $\alpha'=\alpha+1$,
\begin{equation*}
  \begin{split}
  &\Gamma_{\alpha,\alpha+1}(J_\alpha,J_{\alpha+1}) 
  =
  - \frac 1{L^2} \sum_{i\in \mb Z_L}
  \big[ J_\alpha\big(\tfrac{i+1}L\big)-J_\alpha\big(\tfrac iL\big)\big]  
  \big[ J_{\alpha+1}\big(\tfrac{i+1}L\big)-J_{\alpha+1}\big(\tfrac iL\big)\big]  
  \\
  &\quad \times
  \big\{ \mathrm{e}^{-\frac \beta{2L}} \sigma_\alpha(i) \sigma_{\alpha+1}(i+1) + \mathrm{e}^{\frac \beta{2L}}
  \sigma_{\alpha+1}(i) \sigma_{\alpha}(i+1)
  \big\}
  \\ 
  &\; 
  \approx
  - \frac 1{L^2} \frac 1{L^2} \sum_{i\in \mb Z_L}
  J_\alpha'\big( \tfrac iL\big) J_{\alpha+1}'\big( \tfrac iL\big)
  \big\{ \sigma_\alpha(i) \sigma_{\alpha+1}(i+1) +
  \sigma_{\alpha+1}(i) \sigma_{\alpha}(i+1)
  \big\},
  \end{split}
\end{equation*}
and finally, for  $\alpha'=\alpha+2$,
\begin{equation*}
  \begin{split}
  &\Gamma_{\alpha,\alpha+2}(J_\alpha,J_{\alpha+2}) 
  =
  - \frac 1{L^2} \sum_{i\in \mb Z_L}
  \big[ J_\alpha\big(\tfrac{i+1}L\big)-J_\alpha\big(\tfrac iL\big)\big]  
  \big[ J_{\alpha+2}\big(\tfrac{i+1}L\big)-J_{\alpha+2}\big(\tfrac iL\big)\big]  
  \\
  &\quad \times
  \big\{ \mathrm{e}^{\frac \beta{2L}}\sigma_\alpha(i) \sigma_{\alpha+2}(i+1) +
  \mathrm{e}^{-\frac \beta{2L}} \sigma_{\alpha+2}(i) \sigma_{\alpha}(i+1)
  \big\}
  \\ 
  &\; 
  \approx
  - \frac 1{L^2} \frac 1{L^2} \sum_{i\in \mb Z_L}
  J_\alpha'\big( \tfrac iL\big) J_{\alpha+2}'\big( \tfrac iL\big)
  \big\{ \sigma_\alpha(i) \sigma_{\alpha+2}(i+1) +
  \sigma_{\alpha+2}(i) \sigma_{\alpha}(i+1)
  \big\}.
  \end{split}
\end{equation*}
In view of the factorization assumption discussed before, the
fluctuations of the empirical densities agree with the correlation of
the noise in \eqref{fhy}. Compare in particular \eqref{cov0} and 
\eqref{p1.5-1} to the above expressions.

\section{Expansions near the critical point}
\label{app:c}

\subsection{The steady state close to the second order phase transition}
\label{app:c1}

Recalling $\beta_r=2\pi/\sqrt{1-2r^2}$, $\theta=(\beta-\beta_r)/\beta_r$, and setting $\upbar\rho_\alpha=r_\alpha+\Psi_\alpha$, $\alpha\in\{A,B\}$, the stationary equation \eqref{ss} reads,
\begin{equation}
\label{ss1}
\mc A^\beta\Psi + \beta Q(\Psi,\Psi) = 0,
\end{equation}
where
\[
\mc A^\beta f = f'' + \beta\begin{pmatrix} 2r_A+2r_B-1 & 2r_A \\ -2r_B & -2r_A-2r_B+1\end{pmatrix} f'
\]
and
\[
Q(u,v) = \begin{pmatrix} [u_A(v_A+2v_B)]' \\ - [v_B(u_B +2u_A)]'\end{pmatrix}.
\]
It turns out that the solution to \eqref{ss1} can be expanded in powers of $\theta^{1/2}$ and the first order term is explicitly given in \cite{8}. Here we compute also the second order. 

We regard $\mc A^\beta$ as an operator on the space of mean zero functions $f\colon \bb T \to \bb R^2$. The spectrum of $\mc A^\beta$ is easily computed in Fourier basis, and its eigenvalues are given by
\begin{equation}
\label{ln}
\lambda^+_n = -4\pi^2n(n+1+\theta), \qquad \lambda^-_n = -4\pi^2n(n-1-\theta),\qquad n\ge 1,
\end{equation} 
each one having multiplicity two. For $\theta\in (0,1)$ the unique positive eigenvalue is $\lambda^-_1=4\pi^2\theta$. The associated family of right and left eigenvectors is
\[
e_z(x) =\Lambda\, \mathrm{e}^{2\pi \mathrm{i}(x- z)} + \mathrm{c.c.}, \qquad w_z(x) = \Gamma^\top\, \mathrm{e}^{2\pi\mathrm{i} (x-z)} + \mathrm{c.c.}, 
\]
where $z\in [0,1)$, $\Lambda$ is defined in \eqref{LamIup}, and 
\begin{equation}
\label{G}
\Gamma  = \sqrt{\frac{r_B}{r_A}} \mathrm{e}^{-\mathrm{i}\phi} \begin{pmatrix} 0 & 1 \\ -1 & 0 \end{pmatrix} \Lambda = \sqrt{\frac{r_B}{1-2r^2}} \begin{pmatrix} \sqrt{r_B} \\ - \sqrt{r_A} \mathrm{e}^{-\mathrm{i}\phi} \end{pmatrix},
\end{equation}
in which we recall the angle $\phi$ is defined in \eqref{phi}. Observe the above family can be obtained by taking the linear span of two linearly independent vectors. Moreover, they are normalized so that $\lan w_z| e_z \ran=1$, where $\lan u|v\ran = \int\!\mathrm{d} y \: u(y)v(y)$ is the pairing introduced in Section \ref{s:cb}.

Writing $\Psi = \xi e_z+\Psi^1$ with 
\[
\xi = \lan w_z | \Psi \ran, \qquad \lan w_z | \Psi^1\ran= 0,
\]
the stationary equation becomes,
\[
4\pi^2\theta \xi e_z+\mc A^\beta \Psi^1+\xi^2 \beta Q(e_z,e_z) + \xi \beta [Q(e_z,\Psi^1)+Q(\Psi^1,e_z)] + \beta Q(\Psi^1,\Psi^1) = 0.
\]
We observe that
\[
Q(e_z,e_z)  = E  \, \mathrm{e}^{4\pi \mathrm{i}(x- z)} + \mathrm{c.c.}
\]
with
\[
E = 4\pi \mathrm{i} \begin{pmatrix}\Lambda_A(\Lambda_A+2\Lambda_B)\\ -\Lambda_B(\Lambda_B+2\Lambda_A)\end{pmatrix}.
\]
In particular,
\[
\lan w_z | Q(e_z,e_z)\ran = 0.
\]
As also
\[
\lan w_z | \mc A^\beta\Psi^1\ran = 4\pi^2\theta \lan w_z | \Psi^1\ran = 0,
\]
the stationary equation can be recast in the form,
\[
\begin{cases} 4\pi^2\theta \xi + \xi\beta \lan w_z | Q(e_z,\Psi^1)+Q(\Psi^1,e_z) \ran+ \beta \lan w_z | Q(\Psi^1,\Psi^1)\ran  = 0, \\ \mc A^\beta \Psi^1+\xi^2 \beta Q(e_z,e_z) + \xi \beta [Q(e_z,\Psi^1)+Q(\Psi^1,e_z)] + \beta Q(\Psi^1,\Psi^1) = 0. \end{cases}
\]
Assuming $\Psi^1\to 0$ for $\theta\downarrow 0$, at the lower order in $\theta$ the above system reads,
\[
\begin{cases} 4\pi^2\theta \xi + \xi\beta_r \lan w_z | Q(e_z,\Psi^1)+Q(\Psi^1,e_z) \ran = 0, \\ \mc A^{\beta_r} \Psi^1+\xi^2 \beta_r Q(e_z,e_z)  = 0. \end{cases}
\]
The second equation imposes that $\Psi^1 = M \, \mathrm{e}^{4\pi \mathrm{i}(x- z)} + \mathrm{c.c.}$ with $M$ solution to 
\[
\begin{cases} [-16\pi^2+4\pi\mathrm{i} \beta_r(2r_A+2r_B-1)] M_A + 8\pi\mathrm{i} \beta_r r_A M_B = - \xi^2 \beta_r E_A, \\ - 8\pi\mathrm{i} \beta_r r_B M_A - [16\pi^2+4\pi\mathrm{i} \beta_r(2r_A+2r_B-1)] M_B = - \xi^2 \beta_r E_B, \end{cases}
\]
which can be solved, yielding
\[
M = \frac{\xi^2r_A}{(1-2r^2)^2} \begin{pmatrix} r_A(1-2r_A) \\ r_B(1-2r_B) \mathrm{e}^{2\mathrm{i}\phi} \end{pmatrix}.
\]
Now, it is easy to check that
\[
\begin{split}
K & :=\lan w_z | Q(e_z,\Psi^1)  +Q(\Psi^1,e_z) \ran  = 4\pi\mathrm{i} (\upbar\Lambda_A M_A + \upbar\Lambda_A M_B + \upbar\Lambda_B M_A)\upbar\Gamma_A \\  & \qquad\qquad - 4\pi\mathrm{i} (\upbar\Lambda_B M_B + \upbar\Lambda_B M_A + \upbar\Lambda_A M_B)\upbar\Gamma_B + \mathrm{c.c.}
\end{split}
\]
and therefore, by inserting the explicit expressions of $\Lambda,\Gamma,M$,
\[
\beta_r K = \frac{8\pi^2\xi^2r_A}{(1-2r^2)^3} = \frac{8\pi^2\xi^2r_A}{(1-2r^2)^3} \big[2(r_A^3+r_B^3+r_C^3)-r^2\big],
\]
where we used the identity,
\begin{equation}
\label{id}
4r_A^2+4r_B^2+1-4r_A-4r_B +10r_Ar_B -6r_A^2r_B-6r_Ar_B^2 = 2(r_A^3+r_B^3+r_C^3)-r^2,
\end{equation}
which holds whenever $r_C=1-r_A-r_B$. Plugging this expression in the scalar equation $4\pi^2\theta \xi + \xi\beta_r K = 0$, we obtain the not zero solutions,
\[
\xi = \pm \frac{(1-2r^2)^{3/2}\theta^{1/2}}{\sqrt{r_A}\sqrt{2r^2-4(r_A^3+r_B^3+r_C^3)}} = \pm \psi(\theta)\sqrt{\frac{1-2r^2}{r_A}},
\]
where $\psi(\theta)$ is defined in \eqref{tp}. It is sufficient to consider only the positive solution, as the opposite one gives rise to the same profile shifted by one half (i.e., with $z+\frac 12$ instead of $z$). In conclusion, neglecting terms of order $\theta^{3/2}$,
\[
\Psi(x) = \psi(\theta) \begin{pmatrix} \sqrt{r_A} \\ \sqrt{r_B} \mathrm{e}^{\mathrm{i}\phi} \end{pmatrix} \mathrm{e}^{2\pi\mathrm{i} (x-z)} + \frac{\psi(\theta)^2}{1-2r^2} \begin{pmatrix} r_A(1-2r_A) \\ r_B(1-2r_B) \mathrm{e}^{2\mathrm{i}\phi} \end{pmatrix}\mathrm{e}^{4\pi\mathrm{i} (x-z)} + \mathrm{c.c.} 
\]
which proves \eqref{r=}.

\subsection{Linear perturbation theory}
\label{app:c2}

Here we compute the perturbative expansion, as $\theta\downarrow 0$, of the spectral gap of $\mc L_z$ defined in \eqref{p2} and of its associated eigenvectors. Without loss of generality, 
from now on we fix $z=0$ and drop the subscript. As
\[
\mc L f= f'' + (1+\theta) (\mc V  f)', \qquad \mc V = \beta_r\begin{pmatrix} 2\upbar{\rho}_A+2\upbar{\rho}_B-1 & 2\upbar{\rho}_A \\ -2\upbar{\rho}_B & -2\upbar{\rho}_A-2\upbar{\rho}_B+1 \end{pmatrix},
\]
we have,
\[
\mc L f = \mc A^{\beta_r} f + (\mc V^1f)' + \theta\mc V^0 f' + (\mc V^2f)' + O(\theta^{3/2}),
\]
with
\[
\mc V^0 = \frac{2\pi}{\sqrt{1-2r^2}}\begin{pmatrix} 2r_A+2r_B-1 & 2r_A \\ -2r_B & -2r_A-2r_B+1\end{pmatrix}
\]
and 
\[
\mc V^1 = \frac{ 4\pi\psi(\theta)}{\sqrt{r_A(1-2r^2)}} V^{(1)} \mathrm{e}^{2\pi \mathrm{i}x} +  \mathrm{c.c.}, \qquad \mc V^2 =  \frac{ 4\pi\psi(\theta)^2}{(1-2r^2)^{3/2}} V^{(2)}\mathrm{e}^{4\pi \mathrm{i}x} +  \mathrm{c.c.},
\]
where 
\[
V^{(1)} = \begin{pmatrix} r_A+\sqrt{r_Ar_B}\,  \mathrm{e}^{\mathrm{i}\phi} & r_A  \\ - \sqrt{r_Ar_B}\, \mathrm{e}^{\mathrm{i}\phi} & - r_A - \sqrt{r_Ar_B}\, \mathrm{e}^{\mathrm{i}\phi} \end{pmatrix}
\]
and 
\[
V^{(2)} = \begin{pmatrix} r_A(1-2r_A) + r_B(1-2r_B) \mathrm{e}^{2\mathrm{i}\phi} & r_A(1-2r_A)  \\ - r_B(1-2r_B) \mathrm{e}^{2\mathrm{i}\phi} & - r_A(1-2r_A) -  r_B(1-2r_B) \mathrm{e}^{2\mathrm{i}\phi}  \end{pmatrix}.
\]

For $\theta=0$ the operator $\mc L$ coincides with $\mc A^{\beta_r}$. As follows from \eqref{ln}, zero is an eigenvalue of $\mc A^{\beta_r}$ with multiplicity two.  Accordingly, perturbation theory of linear operators implies that for $\theta$ small $\mc L$ has two small eigenvalues that can be computed perturbatively. Since we already know that for $\theta>0$ the operator $\mc L$ has zero as a simple eigenvalue, we can compute the asymptotic expansion of the other (necessarily real) eigenvalue without really using degenerate perturbation theory. 

The right and left eigenvectors of $\mc A^{\beta_r}$ associated to the zero eigenvalue and bi-orthonormal (up to the order $O(\theta)$) to $\upbar\rho'$ and $\chi$ are 
\[
e_0(x) =\Lambda\, \mathrm{e}^{2\pi \mathrm{i}x} + \mathrm{c.c.}, \qquad w_0(x) = \sqrt{\frac{1-2r^2}{4r_Ar_B}} \, \mathrm{i}\,\mathrm{e}^{\mathrm{i}\phi} \Gamma^\top\, \mathrm{e}^{2\pi\mathrm{i} x} + \mathrm{c.c.},
\]
where $\Lambda$ and $\Gamma$ are defined in \eqref{LamIup} and \eqref{G}. Let $\lambda=\lambda(\theta)$ be the small eigenvalue of $\mc L$ and denote by $e(x)$ and $w(x)$ its right and left eigenvector. We expand $\lambda=\lambda_1+\lambda_2 + O(\theta^{3/2})$ in powers of $\theta^{1/2}$ and, accordingly,
\[
e(x) = e_0(x) + e_1(x) + e_2(x) + O(\theta^{3/2}).
\]
Plugging these expansions in the eigenvalue equation $\mc L e=\lambda e$ we get,
\[
\begin{split}
\mc A^{\beta_r} e_1 + (\mc V^1e_0)' & = \lambda_1e_0, \\  \mc A^{\beta_r} e_2 + (\mc V^1 e_1)' + \theta \mc V^0 e_0' + (\mc V^2e_0)' & = \lambda_1e_1+\lambda_2e_0.
\end{split}
\]
As $\langle w_0 | \mc A^{\beta_r}e_1\rangle = \langle w_0 | \mc A^{\beta_r}e_2\rangle = 0$, $\langle w_0 | e_0\rangle = 1$,  and $\langle w_0 | \mc V^0 e_0'\rangle = 4\pi^2$, projecting onto $w_0$ gives,
\[
\begin{split}
\langle w_0 |  (\mc V^1e_0)'\rangle & = \lambda_1, \\
\langle w_0 | (\mc V^1 e_1)' \rangle + 4\pi^2 \theta + \langle w_0 | (\mc V^2e_0)' \rangle & = \lambda_1 \langle w_0 | e_1 \rangle +\lambda_2.
\end{split}
\]
By an explicit computation, $(\mc V^1e_0)' = C \mathrm{e}^{4\pi\mathrm{i} x} + \mathrm{c.c.}$ with
\[
C = \frac{16\pi^2\mathrm{i} \sqrt{r_A}\psi(\theta)}{1-2r^2} \begin{pmatrix}r_A+2\sqrt{r_Ar_B}\, \mathrm{e}^{\mathrm{i}\phi} \\ - 2\sqrt{r_Ar_B}\, \mathrm{e}^{\mathrm{i}\phi} - r_B  \, \mathrm{e}^{2\mathrm{i}\phi} \end{pmatrix}.
\]
It follows that $\lambda_1=0$ and that $e_1 = \psi(\theta) \Upsilon\, \mathrm{e}^{4\pi\mathrm{i} x} + \mathrm{c.c.}$ with $\Upsilon$ solution  to
\[
\begin{cases} [-16\pi^2+4\pi\mathrm{i} \beta_r(2r_A+2r_B-1)] \Upsilon_A + 8\pi\mathrm{i} \beta_r r_A \Upsilon_B = - \psi(\theta)^{-1}C_A, \\ - 8\pi\mathrm{i} \beta_r r_B \Upsilon_A - [16\pi^2+4\pi\mathrm{i} \beta_r(2r_A+2r_B-1)] \Upsilon_B = - \psi(\theta)^{-1}C_B, \end{cases}
\]
which can be solved, yielding $\Upsilon$ as in \eqref{LamIup}. Hence, $\lambda=\lambda_2+ O(\theta^{3/2})$ with
\[
\lambda_2 = 4\pi^2 \theta + \langle w_0 | (\mc V^1 e_1)' + (\mc V^2e_0)' \rangle.
\]
It remains to compute the scalar product in the right-hand side. We have, for a suitable $W$,
\[
(\mc V^1 e_1)'  + (\mc V^2 e_0)'  = \frac{24\pi^2\mathrm{i}\psi(\theta)^2}{(1-2r^2)^2} S\mathrm{e}^{2\pi\mathrm{i} x}  +  W \mathrm{e}^{6\pi\mathrm{i} x} + \mathrm{c.c.}, 
\]
where
\[
S = \begin{pmatrix} r_A^2(1-2r_A) + r_A(1-2r_A) \sqrt{r_Ar_B}\mathrm{e}^{-\mathrm{i}\phi} +  r_Ar_B(1-2r_B)\mathrm{e}^{2\mathrm{i}\phi} \\  -  r_A(1-2r_A) \sqrt{r_Ar_B}\mathrm{e}^{-\mathrm{i}\phi}-  r_B(1-2r_B) \sqrt{r_Ar_B}\mathrm{e}^{\mathrm{i}\phi} - r_Ar_B(1-2r_B)\mathrm{e}^{2\mathrm{i}\phi} \end{pmatrix}.
\]
Therefore, after some explicit computation and using the identity \eqref{id},
\[
\begin{split}
\langle w_0 | (\mc V^1 e_1)' + (\mc V^2e_0)' \rangle  & = \frac{12\pi^2\psi(\theta)^2}{(1-2r^2)^2} \sqrt{\frac{1-2r^2}{4r_Ar_B}} \, \mathrm{e}^{-\mathrm{i}\phi} \, \upbar\Gamma^\top S + \mathrm{c.c.}  \\ & = \frac{24\pi^2\psi(\theta)^2}{(1-2r^2)^2} \big[2(r_A^3+r_B^3+r_C^3)-r^2\big] .
\end{split}
\]
Recalling the definition of $\psi(\theta)$ in \eqref{tp}, we finally get,
\[
\lambda_2 = - 8\pi^2 \theta. 
\]

An evocative explanation of this result is the following. Consider the one dimensional equation $F_\theta(x) = 0$ with $F_\theta(x) = \theta x - \alpha x^3$, $\alpha>0$. When $\theta$ crosses the threshold zero, a pitchfork bifurcation takes place and the stable solutions are $x_\pm(\theta) = \pm \sqrt{\theta/\alpha}$. Then, irrespectively of $\alpha$, $F_\theta'(x_\pm(\theta)) = -2F_\theta'(0)$. Observe in fact that the stationary equation \eqref{ss1} in the  direction $e_z$ has an analogous structure near the bifurcation point. 

\subsection*{Acknowledgments}

We are grateful to T.\ Bodineau, B.\ Derrida, and C.\ Mascia for interesting exchanges of views. We also thank an anonymous referee for suggesting us the analysis near the critical point and for other valuable comments.

\end{document}